\documentstyle[12pt]{article}
\def\beq{\begin{equation}}
\def\eeq{\end{equation}}
\def\bea{\begin{eqnarray}}
\def\eea{\end{eqnarray}}
\def\bq{\begin{quote}}
\def\eq{\end{quote}}

\def\gappeq{\mathrel{\rlap {\raise.5ex\hbox{$>$}}
{\lower.5ex\hbox{$\sim$}}}}

\def\lappeq{\mathrel{\rlap{\raise.5ex\hbox{$<$}}
{\lower.5ex\hbox{$\sim$}}}}

\begin{document}
\pagestyle{empty}
\begin{flushright}
{CERN-TH/96-110}
\end{flushright}
\vspace*{5mm}
\begin{center}
{\bf QUARK MASS EFFECTS IN \\
DEEPLY INELASTIC SCATTERING} \\
\vspace*{1cm} 
{\bf A.V. Kisselev}$^{*)}$ \\
\vspace{0.3cm}
Theoretical Physics Division, CERN \\
CH - 1211 Geneva 23 \\
and \\
Institute for High Energy Physics \\
142284 Protvino, Russia \\
\vspace*{0.3cm}
{\bf V.A. Petrov} \\
\vspace{0.3cm}
Institute for High Energy Physics \\
142284 Protvino, Russia \\
\vspace*{2cm}  
{\bf Abstract} \\ \end{center}
\vspace*{5mm}
\noindent
\begin{quote}
We argue that the difference between the structure functions corresponding
to deep inelastic scattering with and without heavy quarks in the current
fragmentation region scales at high $Q^2$ and fixed (low) $x_{Bj}$.
\end{quote}

\vspace*{3cm} 
\begin{flushleft} CERN-TH/96-110 \\
April 1996
\end{flushleft}
\vspace*{0.5cm}

\noindent
\rule[.1in]{17cm}{.002in}

\noindent
$^{*)}$ Permanent address: Institute for High Energy Physics,
142284 Protvino, Russia. E-mail address: kisselev@mx.ihep.su. 

\vfill\eject

\setcounter{page}{1}
\pagestyle{plain}
%
%
\section{Introduction}

Quite often mass effects in high--energy collisions are considered as some not 
very spectacular corrections which finally die off. Nonetheless, it appears 
that in $e^+e^-$ annihilation even such overall characteristics as hadron
multiplicities are quite sensitive to the value of masses of the primary
$q \bar q$ pairs~\cite{Data}. 

Recent considerations have shown that 
calculations based on QCD agree well with the data at high enough 
energy~\cite{Petrov-95} and that they yield an asymptotically constant 
difference between 
multiplicities of hadrons induced by the primary quarks of different masses.

In this paper we consider the possibility of a similar effect in a deeply 
inelastic process.

\section{Calculation of quark mass dependence}
Let us consider, for definiteness, deep inelastic scattering of the electron
(muon) off the proton. The hadronic tensor (an imaginary part of the virtual 
photon--proton amplitude) is defined via the electromagnetic current $J_{\mu}$:
\begin{equation}
W_{\mu \nu} (p,q) = \frac{1}{2} (2\pi)^2 \int d^4z \exp (iqz)
\langle p|[J_{\mu}(z), J_{\nu}(0)]|p \rangle, \label{1}
\end{equation}
where $p$ is the momentum of the proton, $p^2=M^2$, and $q$ is the momentum of
virtual photon, $q^2=-Q^2<0$.

A symmetric part of $W_{\mu \nu}$ has two Lorentz structures:
\begin{equation}
W_{\mu \nu} = \left(- g_{\mu \nu} + \frac{q_{\nu}q_{\nu}}{q^2} 
\right) F_1(Q^2,x) 
+ \frac{1}{pq} \left( p_{\mu} - q_{\mu} \frac{pq}{q^2}\right) 
\left( p_{\nu} - q_{\nu} \frac{pq}{q^2}\right) F_2(Q^2,x), \label{2}
\end{equation}
where the structure functions $F_1$ and $F_2$ depend on $Q^2$ and on the 
variable
\begin{equation}
x = \frac{Q^2}{pq + \sqrt{(pq)^2 + Q^2 M^2}}\raise 2pt \hbox{.} \label{3}
\end{equation}

In what follows we will analyse the structure function $F_2$
of deep inelastic scattering with open charm (beauty) production at small $x$. 
In this section we consider the case of one single quark loop with mass $m_q$
and electric charge $e_q$. A general case will be discussed in Section~2.

At small $x$ a leading contribution to $F_2$ comes from one photon--gluon
fusion subprocess~\cite{Catani-91}:
\begin{equation}
W_{\mu \nu} = \int \frac{d^4k}{(2\pi)^4} \frac{1}{k^4}
C_{\mu \nu}^{\alpha \beta} (q,k;m_q)
 d_{\alpha \alpha'}(k) d_{\beta \beta'}(k)
\Gamma^{\alpha' \beta'} (k,p), \label{4}
\end{equation}
where $k$ is the momentum of the virtual gluon, $k^2<0$. The tensor 
$C_{\mu \nu}^{\alpha \beta}$ denotes an imaginary two gluon irreducible 
part of the photon--gluon amplitude, while $\Gamma^{\alpha \beta}$ describes a 
distribution of the gluon inside the proton. A quantity $d_{\alpha \beta}$
is a tensor part of the gluonic propagator.

Let us choose an infinite momentum frame 
\begin{equation}
p_{\mu} = \biggl(P + \frac{M^2}{4P},0,0, P - \frac{M^2}{4P}\biggr) \raise 2pt
\hbox{.} \label{5}
\end{equation}
Then the gluon distribution 
$\Gamma^{\alpha \beta}$ has to be calculated in the axial gauge $nA=0$ with
a gauge vector $n_{\mu}=(1,0,0,-1)$~\cite{Catani-91}. One can take, 
for instance,
\begin{equation}
n_{\mu} = q_{\mu} + x p_{\mu} \label{6}
\end{equation}
with $x$ defined by Eq.~(\ref{3}).

From Eq.~(\ref{2}) we get
\begin{equation}
\frac{1}{x} F_2 = \biggl[ -g_{\mu \nu} + p_{\mu} p_{\nu}
\frac{3Q^2}{(pq)^2 + Q^2 M^2} \biggl] W^{\mu \nu} \equiv F_2^{(a)} 
+ F_2^{(b)}. \label{7}
\end{equation}
Two terms in the RHS of Eq.~(\ref{7}), $F_2^{(a)}$ and $F_2^{(b)}$, 
correspond to two tensor projectors, $g_{\mu \nu}$ and $p_{\mu}p_{\nu}$.

Note that the structure function $F_L=F_2-2xF_1$ is completely defined 
by the term  $p_{\mu}p_{\nu}$ and, thus, proportional to $F_2^{(b)}$.

By definition, the gluon distribution $\Gamma^{\alpha \beta}$ can be rewritten
in the form
\begin{equation}
\Gamma^{\alpha \beta} = \frac{1}{4\pi} \sum_n \delta ( p + k - p_n) 
\langle p|I_{\alpha}^g(0)|n> <n|I_{\beta}^g(0)|p \rangle, \label{8}
\end{equation}
where $I_{\alpha}^g$ is the conserved current. Both $|p \rangle$ and 
$|n \rangle$ are on shell states that result in
\begin{equation}
k^{\alpha} \Gamma_{\alpha \beta} = 0. \label{9}
\end{equation}

From an explicit form for $C_{\mu \nu}^{\alpha \beta}$ (see Appendix~I) one 
can verify that it obeys the same condition:
\begin{equation}
k^{\alpha} C_{\alpha \beta}^{\mu \nu} = 0. \label{10}
\end{equation}

Equations~(\ref{9}) and (\ref{10}) allow us to simplify expression~(\ref{4}) 
and get ($r=a,b$):
\begin{equation}
\frac{1}{x} F_2^{(r)} = \int \frac{d^4k}{(2\pi)^4} \frac{1}{k^4}
C_{\alpha \beta}^{(r)} (q,k;m_q) \Gamma^{\alpha \beta} (k,p), \label{11}
\end{equation}
with the notations
\begin{eqnarray}
C_{\alpha \beta}^{(a)} &=& - g_{\mu \nu} C_{\alpha \beta}^{\mu \nu},
\nonumber \\
C_{\alpha \beta}^{(b)} &=&  \frac{3Q^2}{(pq)^2 + Q^2 M^2} p_{\mu} p_{\nu} 
C_{\alpha \beta}^{\mu \nu}. \label{12}
\end{eqnarray}

The tensor $\Gamma^{\alpha \beta}$ can be expanded in Lorentz structures
\begin{eqnarray}
\Gamma^{\alpha \beta} &=& \biggl(g_{\alpha \beta} - \frac{k_{\alpha} 
k_{\beta}}{k^2} \biggl) \Gamma_1 + \biggl(p_{\alpha} - k_{\alpha}
\frac{pk}{k^2} \biggl)\biggl(p_{\beta} - k_{\beta} \frac{pk}{k^2} \biggl)
\frac{1}{k^2} \Gamma_2 \nonumber \\
&+& \biggl(k_{\alpha} - n_{\alpha} \frac{k^2}{kn} \biggl)
\biggl(k_{\beta} - n_{\beta} \frac{k^2}{kn} \biggl) \frac{1}{k^2} \Gamma_3 
+ \biggl(p_{\alpha} - n_{\alpha} \frac{pk}{kn} \biggl)
\biggl(p_{\beta} - n_{\beta} \frac{pk}{kn} \biggl) \frac{1}{k^2}
\Gamma_4 \label{13}
\end{eqnarray}
with $\Gamma_i=\Gamma_i (k^2,M^2,pk)$.

Let us consider a contribution of the invariant function $\Gamma_1$ into the 
structure function $F_2$~(\ref{11}). With the accounting for (\ref{9})
and (\ref{10}) we obtain
\begin{equation}
\frac{1}{x} F_2^{(r)} = e_q^2 \int \limits_x^1 \frac{dz}{z}
\int \limits_{Q_0^2}^{Q^2(z/x)} \frac{dl^2}{l^2}
\frac{1 - l^2 x^2/Q^2 z^2}{1 + M^2 x^2/Q^2} C^{(r)} \left( 
\frac{Q^2}{l^2}, \frac{m_q^2}{l^2}, \frac{x}{z} \right) 
\frac{\partial}{\partial \ln l^2} g(l^2,z), \label{14}
\end{equation}
where
\begin{equation}
l^2 = - k^2 > 0, \label{15}
\end{equation}
\begin{equation}
z = \frac{kn}{pn} \label{16}
\end{equation}
and
\begin{equation}
Q_0^2 = \frac{M^2 z^2}{1 - z}. \label{17}
\end{equation}
Here we used the notation:
\begin{equation}
C^{(r)} = - g^{\alpha \beta} C_{\alpha \beta}^{(r)}. \label{18}
\end{equation}

In Eq.~(\ref{14}) the gluon distribution, $g(l^2,z)$, is introduced:
\begin{equation}
g(l^2,z) = \frac{1}{2(2\pi)^4} \int \limits_{Q_0^2}^{l^2} \frac{dl'^2}{l'^4}
\int d^2 k_{\bot} \Gamma_1 (l'^2,k_{\bot},z). \label{19}
\end{equation}
If we use the new variable
\begin{equation}
\xi = \frac{-k^2}{pk + \sqrt{(pk)^2 - k^2 M^2}} \label{20}
\end{equation}
instead of $k_{\bot}^2$, we will arrive at the expression
\begin{equation}
g(l^2,z) = \frac{z}{32\pi^3} \int \limits_{Q_0^2}^{l^2} \frac{dl'^2}{l'^4}
\int \limits_z^1 d\xi \left( M^2 + \frac{l'^2}{\xi^2} \right)
\Gamma_1 (l'^2,\xi). \label{21}
\end{equation}

A thorough analysis shows, however, that the main contribution to $F_2$ 
at small $x$ comes from $\Gamma_2$ and $\Gamma_4$ in~(\ref{13}) (see also
Appendix~II). In Appendix~I the following formula for $F_2$ is obtained:
\begin{eqnarray}
\frac{1}{x} F_2 = e_q^2 \sum_{r=a,b} \int \limits_z^1 \frac{dz}{z}
\int \limits_{Q_0^2}^{Q^2(z-x)/x} \frac{dl^2}{l^2} \!\!\! 
&& \biggl[ \tilde C^{(r)} \left( \frac{Q^2}{l^2}, \frac{m_q^2}{l^2}, 
\frac{x}{z} \right) \frac{\partial}{\partial \ln l^2} G(l^2,z) 
\nonumber \\
&& + \ \hat C^{(r)} \left( \frac{Q^2}{l^2}, \frac{m_q^2}{l^2}, 
\frac{x}{z} \right) \frac{\partial}{\partial \ln l^2} \hat G(l^2,z)  
\biggr]. \label{22}
\end{eqnarray}

As we are interested in a calculation of the difference of the structure 
functions corresponding to the massive and massless cases, we preserve
those terms in $C^{(r)}$ which give a leading contribution to $\Delta F_2$.
In Appendix~I we have calculated the functions $C^{(a)}$ 
in lowest order in the strong coupling $\alpha_s$:
\begin{eqnarray}
\tilde C^{(a)} (u,v,y) &=&  \frac{\alpha_s}{4\pi} \{ [(1 - y)^2 + y^2] L(u,v,y) 
- [(1 - y)^2 + y^2 - 2v] M(v,y) - 1\}, \nonumber \\
\hat C^{(a)} (u,v,y) &=& \frac{\alpha_s}{\pi} y(1 - y) M(v,y) , \label{23}
\end{eqnarray}
where
\begin{eqnarray}
L(u,v,y) &=& \ln \frac{u(1-y)}{y [v + y(1 - y)]} \raise 2pt \hbox{,}
\nonumber \\
M(v,y) &=& \frac{y(1-y)}{v + y(1 - y)} \raise 2pt \hbox{.} \label{24}
\end{eqnarray}

As for the gluon distributions, they are given by the formulae:
\begin{eqnarray}
G &=& \frac{1}{32\pi^3z} \int \limits_{Q_0^2}^{l^2} \frac{dl'^2}{l'^4}
\int \limits_z^1 \frac{d\xi}{\xi} (\xi - z) \left( M^2 + \frac{l'^2}{\xi^2} 
\right)
[\Gamma_2 (l'^2,\xi) + \Gamma_4 (l'^2,\xi)], \label{25a} \\
\hat G &=& \frac{1}{32\pi^3z} \int \limits_{Q_0^2}^{l^2} \frac{dl'^2}{l'^4}
\int \limits_z^1 d\xi 
\left( M^2 + \frac{l'^2}{\xi^2} \right) 
\left[ \frac{(2\xi - z)^2}{4\xi^2} 
\Gamma_2 (l'^2,\xi) + \Gamma_4 (l'^2,\xi) \right]. \label{25b}
\end{eqnarray}

The analogous expressions for the functions $C^{(b)}$ are 
the following:
\begin{eqnarray}
\tilde C^{(b)} (u,v,y) &=& \frac{3\alpha_s}{2\pi}
\frac{1}{u} y \{ 2y [(1 - 2y) (1 - y) -v] L(u,v,y) \nonumber \\
&+& (1 - y) [(1 - y)^2 + y^2 - 2v] M(v,y) \} 
+ \frac{3\alpha}{2\pi} y(1 - y), \nonumber \\
\hat C^{(b)} (u,v,y) &=& - \frac{12\alpha_s}{\pi}
\frac{1}{u} y^2 (1 - y)^2 M(v,y). \label{26}
\end{eqnarray}

It may be shown that the leading contribution to $\Delta F_2$ comes from the 
region $l^2 \sim m_q^2$, $k^2=-l^2$ being the gluon virtuality 
(see Appendix~I). 
Then one can easily see from (\ref{24}) and (\ref{27}) that the first two 
terms in 
$\tilde C^{(b)}$ are suppressed  by the factor $k^2/Q^2$ with respect to 
$\tilde C^{(a)}$,
while the third terms in $\tilde C^{(b)}$ do not contribute to the difference 
$C^{(b)}|_{m = 0} - C^{(b)}|_{m \neq 0}$. 

In the leading logarithmic approximation (LLA), only the function $L$  
remaines in Eqs.~(\ref{23}), which results in
\begin{equation}
\frac{1}{x} \frac{\partial}{\partial \ln Q^2} F_2(Q^2,x) = 
\frac{\alpha_s}{2\pi} \int \limits_x^1 \frac{dz}{z}P_{qg} \biggl( \frac{x}{z}
\biggr) G(Q^2,z) \raise 2pt \hbox{,} \label{27}
\end{equation}
where $P_{qg}(z)$ is the Altarelli--Parisi splitting function and $G(Q^2,z)$ 
is the gluon distribution in LLA defined by Eq.~(\ref{25a}). 

The gluon distribution~(\ref{25a}) in LLA can be rewritten in the form
(compare it with the corresponding formulae in Appendix~II)
\begin{equation}
G(Q^2,z) =  \frac{1 - z}{4z^2}\int \limits_{Q_0^2}^{Q^2} 
\frac{dl^2}{l^4} \int \frac{d^2 k_\bot}{(2\pi)^4}
[\Gamma_2 (l^2,k_\bot,z) + \Gamma_4 (l^2,k_\bot,z)]. \label{29}
\end{equation}

It is clear from (\ref{22}) that 
$\Delta F_2 = F_2|_{m = 0} - F_2|_{m \neq 0}$ is defined by the 
quantities ($r=a,b$)
\begin{equation}
\Delta C^{(r)}(u,v,y) = C^{(r)}(u,0,y) - C^{(r)}(u,v,y). \label{30}
\end{equation}
By using Eq.~(\ref{23}) we obtain the important result
\begin{eqnarray}
\Delta \tilde C^{(a)} &=& \Delta \tilde C^{(a)}(v,y) \nonumber \\
\Delta \hat C^{(a)} &=& \Delta \hat C^{(a)}(v,y), \label{31}
\end{eqnarray}
while from (\ref{26}) we get
\begin{eqnarray}
\Delta \tilde C^{(b)} &=& \frac{1}{u} \Delta \tilde C^{(b)}(v,y),
\nonumber \\
\Delta \hat C^{(b)} &=& \frac{1}{u} \Delta \hat C^{(b)}(v,y). \label{32}
\end{eqnarray}
In this, we have
\begin{equation}
\Delta \tilde C^{(a)}, \Delta \hat C^{(a)} 
|\mathstrut_{-k^2 \rightarrow \infty} \sim \frac{m_q^2}{k^2}. 
\label{33}
\end{equation}
So, we get~\cite{Kisselev-95}
\begin{eqnarray}
\frac{1}{x} \Delta F_2 (Q^2,m_q^2,x) |\mathstrut_{Q^2 \rightarrow 
\infty} = e_q^2 \int \limits_x^1 \frac{dz}{z}
\int \limits_{Q_0^2}^{\infty} \frac{dl^2}{l^2} \!\!\! 
&& \biggl[ \Delta \tilde C \left( \frac{m_q^2}{l^2},\frac{x}{z} \right) 
\frac{\partial}{\partial \ln l^2} G(l^2,z) \nonumber \\ 
&& + \ \Delta \hat C \left( \frac{m_q^2}{l^2}, \frac{x}{z} \right)
\frac{\partial}{\partial \ln l^2} \hat G(l^2,z) \biggr]. \label{34}
\end{eqnarray}
Here
\begin{eqnarray}
\Delta \tilde C(v,y) &=& \frac{\alpha_s}{4\pi} \biggl\{ [(1 - y)^2 + y^2] 
\ln \biggl[ 1 + {\displaystyle \frac{v}{y(1-y)}} \biggr] 
- {\displaystyle \frac{v}{v + y(1-y)}} \biggr\}, \nonumber \\
\Delta \hat C(v,y) &=& \frac{\alpha_s}{\pi} 
y(1-y) {\displaystyle \frac{v}{v + y(1-y)}} \label{35}
\end{eqnarray}
with $G(l^2,z)$ and $\hat G(l^2,z)$ being defined by Eqs.~(\ref{25a})
and (\ref{25b}). 

The integral in $l^2$ (\ref{34}) converges because of condition~({\ref{33}). 
Contributions from $\Delta \tilde C^{(b)}$ and $\Delta \hat C^{(b)}$ are 
suppressed by the factors $(m^2/Q^2) \ln Q^2$ and can thus be omitted.

Let us consider the gluon distribution $\hat G$~(\ref{25b}). At small $z$ the
leading contribution to $\hat G(l^2,z)$ comes from the region $z \ll \xi$,
and we have
\begin{equation}
\hat G(l^2,z) \simeq G(l^2,z). \label{36}
\end{equation}
Taking expression~(\ref{36}) into account, the structure function 
$F_2$~(\ref{22}) has the following form at low $x$ (with the term of the order
of $k^2/Q^2$ and $m^2/Q^2$ subtracted)
\begin{equation}
\frac{1}{x} F_2 = e_q^2 \int \limits_x^1 \frac{dz}{z} 
\int \limits_{Q_0^2}^{Q^2(z-x)/x} \frac{dl^2}{l^2} C \left( \frac{Q^2}{l^2},
\frac{m_q^2}{l^2}, \frac{x}{z} \right) 
\frac{\partial}{\partial \ln l^2} G(l^2,z), \label{37}
\end{equation}
where
\begin{equation}
C(u,v,y) = \frac{\alpha_s}{4\pi} \{ [(1 - y)^2 + y^2] L(u,v,y)
- [(1 - 3y)^2 - 3y^2 - 2v] M(v,y) - 1 \}. \label{38}
\end{equation}

As for the difference of the structure function, we obtain the following
prediction
\begin{equation}
\frac{1}{x} \Delta F_2 (Q^2, m_q^2, x) 
= \frac{1}{x} \Delta F_2 (m_q^2,x) = e_q^2 \int \limits_x^1 \frac{dz}{z} 
\int \limits_{Q_0^2}^{\infty}
\frac{dl^2}{l^2} \Delta C \left( \frac{m_q^2}{l^2}, \frac{x}{z} \right)
\frac{\partial}{\partial \ln l^2} G(l^2,z), \label{39}
\end{equation}
where
\begin{equation}
\Delta C(v,y) = \frac{\alpha_s}{4\pi} [(1 - y)^2 + y^2] 
\biggl\{ \ln \biggl[1 + {\displaystyle \frac{v}{y(1 - y)}} \biggr] 
- (1 - 2y)^2 {\displaystyle \frac{v}{v + y(1 - y)}} \biggr\}. \label{40}
\end{equation}

\section{Relation between measurable structure functions}

Up to now, we considered those contributions to $F_2$ that came from the
quark with electric charge $e_q$ and mass $m_q$, $\tilde F_2|_{m \neq 0}$.
Then we have taken the analogous contributions from the massless quark with
the same $e_q$, $\tilde F_2|_{m = 0}$, and have calculated the quantity
$\tilde F_2|_{m = 0} - \tilde F_2|_{m \neq 0}$.

The total structure function $F_2$ has the form
\begin{equation}
F_2(Q^2,x) = \sum_q e_q^2 \tilde F_2^q(Q^2,x), \label{41}
\end{equation}
where the functions $\tilde F_2^q$ are introduced ($q = u,d,s,c,b$).

The structure functions describing open charm and bottom production in DIS,
$F_2^c$ and $F_2^b$ respectively, are related to $\tilde F_2^c$ and 
$\tilde F_2^b$ by the formulae
\begin{eqnarray}
F_2^c &=& \frac{4}{9} \tilde F_2^c, \nonumber \\
F_2^b &=& \frac{1}{9} \tilde F_2^b. \label{42}
\end{eqnarray}

At low $x$ one can put ($m_u = m_d = m_s = 0$ is assumed)
\begin{equation}
\tilde F_2^u = \tilde F_2^d = \tilde F_2^s = \tilde F_2 \label{43}
\end{equation}
and define the difference between heavy and light flavour contributions to
$F_2$:
\begin{eqnarray}
\Delta \tilde F_2^c = \tilde F_2 - \tilde F_2^c, \nonumber \\
\Delta \tilde F_2^b = \tilde F_2 - \tilde F_2^b. \label{44}
\end{eqnarray}
Notice that there are the functions $\tilde F_2$ and $\tilde F_2^q$ that 
have been calculated in the previous section (see Eqs~(\ref{37}) 
and (\ref{39})).

Let us now represent the function $\tilde F_2$~(\ref{37}) in the following
form 
\begin{equation}
\frac{1}{x} \tilde F_2 = \int \limits_x^1 \frac{dy}{y} \int \limits_0^Y
d \eta \, C(\eta, y) \frac{\partial}{\partial \ln l^2} G \left(Y - \eta, 
\frac{x}{y} \right), \label{45}
\end{equation}
where we denote
\begin{equation}
Y = \ln \frac{Q^2}{y Q_0^2} \label{46}
\end{equation}
and introduce the variable $\eta = \ln (k^2/Q_0^2)$.

Analogously, we get from (\ref{39})
\begin{equation}
\frac{1}{x} \Delta \tilde F_2^q = \int \limits_x^1 \frac{dy}{y} \int 
\limits_{- \infty}^{Y_m} d \eta \, \Delta C(\eta, y) 
\frac{\partial}{\partial \ln l^2}
G \left( Y_m - \eta, \frac{x}{y} \right), \label{47}
\end{equation}
with
\begin{equation}
Y_m = \ln \frac{m_q^2}{y Q_0^2}. \label{48}
\end{equation}
Here $\eta = \ln (m_q^2/k^2 y(1-y)) \simeq \ln (m_q^2/k^2 y)$ (remember that
we consider small $x$).

The expression for $\Delta C$ is given by Eq.~(\ref{40}) and,  
in terms of the variables $\eta$ and $y$, looks like 
\begin{equation}
\Delta C = \frac{\alpha_s}{4\pi} [(1 - y)^2 + y^2] \biggl[ \ln \left(
1 + e^{\eta} \right) - (1 - 2y)^2 \frac{e^{\eta}}{1 + e^{\eta}} \biggr].
\label{49}
\end{equation}

As for the expression for $C$, it has to be defined via relation~(\ref{11})
and exact formulae~(\ref{I.14}) and (\ref{I.21}) taken at $m = 0$. The result
of our calculations is of the form
\begin{equation}
C(\eta, y) = \frac{\alpha_s}{2\pi} \left[ \frac{1}{2U} \ln \frac{1 + U}{1 - U}
\biggl( 1 - \frac{3}{U^2} V + V \biggr) - \biggl( 1 - \frac{3}{U^2} V \biggr)
\right], \label{50}
\end{equation}
where
\begin{eqnarray}
U &=& \sqrt{1 - 4y (1 - y) e^{-\eta}}, \nonumber \\
V &=& (1 - y) \left[ y + (1 - y) e^{-\eta} \right] 
\left( 1 - e^{-\eta} \right). \label{51}
\end{eqnarray}

It is clear from (\ref{49}) that
\begin{equation}
\Delta C(\eta, y) > 0 \label{52}
\end{equation}
for $- \infty < \eta < \infty$, $0 \leq y \leq 1$ and $\Delta C(\eta, y)$
is negligible at $\eta < 0$ (see Figs.~1a-1d).

Moreover, the quantitative analysis shows that at most at $y \leq 0.2$, 
which is relevant for small $x$ as under consideration, one has
\begin{equation}
C(\eta, y) > \Delta C(\eta, y), \qquad \eta > 0, \label{53}
\end{equation}
(see Figs.~2a-2d).
Neglecting the small contribution to $\tilde F_2$ from the region $\eta < 0$
and taking into account that $\partial G(Q^2, x) / \partial \ln Q^2 > 0$ at
small $x$ (cf.~\cite{Kwiecinski-96}), we thus conclude
\begin{equation}
\Delta \tilde F_2^q (m_q^2, x) < \tilde F_2 (Q^2, x)|_{Q^2 = m_q^2}. \label{54}
\end{equation}

In terms of the observables $F_2$, $F_2^c$ and $F_2^b$, the 
inequalities~(\ref{52}) and (\ref{54}) can be cast in the forms
\begin{eqnarray}
\left( F_2 - 2.5 F_2^c - F_2^b \right) (Q^2,x)|_{{\rm large} \ Q^2} &>& 0,
\nonumber \\
\left( F_2 - F_2^c - 7 F_2^b \right) (Q^2,x)|_{{\rm large} \ Q^2} &>& 0 
\label{55}
\end{eqnarray}
and
\begin{eqnarray}
\left( F_2 - 2.5 F_2^c - F_2^b \right) (Q^2,x) &<& 
\left( F_2 - F_2^c - F_2^b \right) (Q^2,x) |_{Q^2 = m_c^2},
\nonumber \\
\left( F_2 - F_2^c -7 F_2^b \right) (Q^2,x) &<& 
\left( F_2 - F_2^c - F_2^b \right) (Q^2,x)|_{Q^2 = m_b^2}. \label{56}
\end{eqnarray}

Data on the total structure function $F_2$ for $Q^2$ between $1.5 \ 
{\rm GeV}^2$ and $5000 \ {\rm GeV}^2$ and $x$ between $3 \times 10^{-5}$
and $0.32$ are now available~\cite{H1-96}. As for the charm structure 
function, there are preliminary low--$x$ data on $F_2^c$ at $Q^2 = 13 \ 
{\rm GeV}^2$, $23 \ {\rm GeV}^2$ and $50 \ {\rm GeV}^2$ with rather large
errors~\cite{De Roeck-96}. 

Using the first of the inequalities~(\ref{56}) we get (assuming 
$F_2^c (m_c^2, x)$,  $F_2^b (m_c^2, x) \simeq 0$ (cf.~\cite{Martin-94}))
\begin{equation}
F_2^c (Q^2, x) > 0.4 \left[ F_2(Q^2, x) - F_2^b (Q^2, x) - F_2 (m_c^2, x)
\right]. \label{57}
\end{equation}

Let us estimate $F_2^c (Q^2, x)$ from below for $x = 5 \times 10^{-3}$
and $x = 5 \times 10^{-4}$ and several values of $Q^2$. 
There are data on $F_2 (Q^2, x)$ for $Q^2 = 2.5 \ {\rm GeV}^2$, 
$x = 4 \times 10^{-3}$ and $Q^2 = 2.5 \ {\rm GeV}^2$, 
$x= 6.3 \times 10^{-4}$~\cite{H1-96}. The relative bottom contribution,
$F_2^b/F_2$, reaches at most $2$ to $3 \%$ at HERA.

Putting $F_2 (m_c^2, 4.0 \times 10^{-3}) \simeq
F_2 (m_c^2, 5.0 \times 10^{-3})$ and $F_2 (m_c^2, 6.3 \times 10^{-4})
\simeq F_2 (m_c^2, 5.0 \times 10^{-4})$ and choosing $m_c = 1.58$ GeV
we obtain from~(\ref{58}) the lower bounds for $F_2^c$ presented in 
Tables~1 and 2.
\vspace{.5cm}
\begin{center}
\begin{tabular}{||l||c|c|c||} \hline 
$Q^2$, \ ${\rm GeV}^2$ & $12$ & $25$ & $45$ \\
\hline
$F_2^c (Q^2, x)$ & $0.106 \pm 0.054$ & $0.139 \pm 0.044$ & $0.195 \pm 0.046$ \\ 
\hline
\end{tabular}    
\\ 
\vspace{.3cm}
{\bf Table 1:} The lower bounds on $F_2^c (Q^2, x)$ for $x = 5 \times
10^{-3}$.
\end{center}
\vspace{.5cm}
\begin{center}
\begin{tabular}{||l||c|c||} \hline
$Q^2$, \ ${\rm GeV}^2$ & $12$ & $25$ \\
\hline
$F_2^c (Q^2, x)$ & $0.207 \pm 0.063$ & $0.355 \pm 0.070$ \\ 
\hline
\end{tabular}
\\ 
\vspace{.3cm}
{\bf Table 2:} The lower bounds on $F_2^c (Q^2, x)$ for $x = 5 \times
10^{-4}$.
\end{center}
These quantitative estimates of $F_2^c$ do not contradict the preliminary
data on the charm contribution to $F_2$~\cite{De Roeck-96}.
For a detailed comparison of our predictions with the data, an improved 
measurement of the charm component $F_2^c$ is required.

\section*{Conclusions}

In this paper we have demonstrated that the lowest--order quark loop 
contributions to the structure functions at small $x$ contain mass--dependent
terms which scale at high $Q^2$. This effect can be observed experimentally,
and we predict theoretical bounds for the corresponding contributions from
$c$- and $b$-quarks (see Eqs.~(\ref{55}) and (\ref{56}), Tables~1 and 2).

\section*{Acknowledgements}

We are grateful to Profs. G. Veneziano and J.B. Dainton for stimulating
conversations. We also thank Dr. A. De Roeck for a useful discussion of the data
on $F_2^c$ and for sending us the recent data on $F_2$ from the H1 
Collaboration.
One of us (A.V.K.) is indebted to the Theoretical Physics Division of CERN
for its hospitality and support during the course of this work.
\vfill \eject

\def\theequation{I.\arabic{equation}}
\setcounter{equation}{0}

\section*{Appendix I}
In this section we calculate an imaginary part of the photon--gluon 
amplitude averaged in the photon Lorentz indices with the tensors $g_{\mu
\nu}$ and $p_\mu p_\nu$ (see Eqs.~(\ref{12})).

Let us consider the function $C_{\alpha \beta}^{(a)}$. In the first order
in the strong coupling $\alpha_s$ (a one--loop approximation), it can be 
represented in the form
\begin{eqnarray}
C_{\alpha \beta}^{(a)}(Q^2,k^2,qk) &=& \frac{\alpha_s}{\pi^2}
\int d^4r [\frac{1}{l^4} I_{\alpha \beta}^1 (r,q,k;m^2) 
+ \frac{1}{l^2(2qk-l^2)} I_{\alpha \beta}^2 (r,q,k,;m^2)] \nonumber \\
&\times& \delta_+ ((q-r)^2 - m^2) \delta_+ ((k+r)^2 - m^2), \label{I.1}
\end{eqnarray}
where
\begin{eqnarray}
I_{\alpha \beta}^1 = &-& g_{\alpha \beta} \frac{1}{2}
[(l^4 - 2l^2 qk - k^2 Q^2) + 2m^2 k^2] + r_{\alpha} r_{\beta} 2(Q^2 - 2m^2)
\nonumber \\
&+& (r_{\alpha} k_{\beta} + k_{\alpha} r_{\beta}) (Q^2 - 2m^2) 
- (r_{\alpha} q_{\beta} + r_{\alpha} q_{\beta}) l^2 \nonumber \\
&-& (k_{\alpha} q_{\beta} + k_{\alpha} q_{\beta}) l^2 \label{I.2}
\end{eqnarray}
corresponds to a contribution of a ladder diagram, while 
\begin{eqnarray}
I_{\alpha \beta}^2  &=& g_{\alpha \beta} \frac{1}{2}
k^2 [(k^2 - Q^2 + 2qk) - 2m^2] + r_{\alpha} r_{\beta} 2[(Q^2 + k^2)
- 2m^2] \nonumber \\
&-& k_{\alpha} k_{\beta}(l^2 - Q^2) + q_{\alpha} q_{\beta}(l^2 + k^2)
\nonumber \\
&-& (r_{\alpha} k_{\beta} + k_{\alpha} r_{\beta}) [(l^2 - Q^2 -k^2 -qk)
+ 2m^2] \nonumber \\
&-& (r_{\alpha} q_{\beta} + r_{\alpha} q_{\beta}) [(l^2 + Q^2 -k^2 -qk)
- 2m^2] \nonumber \\
&-& (k_{\alpha} q_{\beta} + k_{\alpha} q_{\beta}) \frac{1}{2}
[(Q^2 + k^2) - 2m^2] \label{I.3}
\end{eqnarray}
is a contribution of a crossed one. In Eqs.~(\ref{I.2}) and (\ref{I.3}) we 
denote
\begin{equation}
l^2 = m^2 - r^2. \label{I.4}
\end{equation}

The following equality thus takes place
\begin{eqnarray}
&& \int d^4r r_{\alpha} f(r^2) \delta_+ ((q-r)^2 - m^2) 
\delta_+ ((k+r)^2 - m^2) \nonumber \\
&& = \frac{\pi}{4\sqrt{D}} [k_{\alpha} A_1 (Q^2,k^2,qk;m^2) 
+ q_{\alpha} A_2 (Q^2,k^2,qk;m^2)], \label{I.5}
\end{eqnarray}
where
\begin{eqnarray}
A_1 &=& -\frac{1}{2D} \int \limits_{l_-^2}^{l_+^2} dl^2 [Q^2 (qk + k^2) 
+ l^2 (qk - Q^2)] f(l^2), \nonumber \\
A_2 &=& \frac{1}{2D} \int \limits_{l_-^2}^{l_+^2} dl^2 [l^2 (qk + k^2) 
- k^2 (qk - Q^2)] f(l^2) \label{I.6}
\end{eqnarray}
with
\begin{eqnarray}
l_{\pm}^2 &=& (qk) \pm \biggl[D(1 - \frac{4m^2}{s}) \biggr]^{1/2}, \nonumber \\
D &=& (qk)^2 + Q^2 k^2, \nonumber \\
s &=& k^2 - Q^2 + 2qk. \label{I.7}
\end{eqnarray}

Analogously, we have
\begin{eqnarray}
&& \int d^4r r_{\alpha} r_{\beta} f(r^2) \delta_+ ((q-r)^2 - m^2) 
\delta_+ ((k+r)^2 - m^2) \nonumber \\
&& = \frac{\pi}{4\sqrt{D}} [g_{\alpha \beta} B_1 + k_{\alpha} k_{\beta} B_2 
+ q_{\alpha} q_{\beta} B_3 
+ \frac{1}{2} (k_{\alpha} q_{\beta} + q_{\alpha} k_{\beta}) B_4], 
\label{I.8}
\end{eqnarray}
where
\begin{eqnarray}
B_1 &=& \int \limits_{l_-^2}^{l_+^2} dl^2 \biggl\{ \frac{1}{8D} Rs + 
\frac{1}{2} m^2 \biggr\} f(l^2), \nonumber \\
B_2 &=& \int \limits_{l_-^2}^{l_+^2} dl^2 \biggl\{ -\frac{3}{8D^2} Q^2 Rs  
+ \frac{1}{4D} [(l^2 - Q^2)^2 - 2m^2 Q^2] \biggr\} f(l^2), \nonumber \\
B_3 &=& \int \limits_{l_-^2}^{l_+^2} dl^2 \biggl\{ \frac{3}{8D^2} k^2 Rs  
+ \frac{1}{4D} [(l^2 + k^2)^2 + 2m^2 k^2] \biggr\} f(l^2), \nonumber \\
B_4 &=& \int \limits_{l_-^2}^{l_+^2} dl^2 \biggl\{ -\frac{3}{4D^2} qk Rs 
+ \frac{1}{2D} [R - l^2 s - 2m^2 qk] \biggr\} f(l^2). 
\label{I.9}
\end{eqnarray}
In Eqs.~(\ref{I.9}) a notation
\begin{equation}
R = l^4 - 2l^2 qk - k^2Q^2 \label{I.10}
\end{equation}
is introduced.

Accounting for all that was said above, we get
\begin{equation}
C_{\alpha \beta}^{(a)}(Q^2,k^2,qk) = \frac{\alpha_s}{4\pi} \frac{1}{\sqrt{D}}
\int \limits_{l_-^2}^{l_+^2} dl^2  \biggl[ \frac{1}{l^4} \tilde 
I_{\alpha \beta}^1 (l^2,q,k;m^2) 
+ \frac{1}{l^2(2qk-l^2)} \tilde I_{\alpha \beta}^2 (l^2,q,k,;m^2) \biggr] 
\label{I.11}
\end{equation}
with
\begin{eqnarray}
\tilde I_{\alpha \beta}^1 = &-& \frac{1}{2} \biggl\{ g_{\alpha \beta} 
\biggl[1 - \frac{1}{2D} Q^2 s \biggr] R - k_{\alpha} k_{\beta} \frac{Q^2}{D} 
\biggl[ R - Q^2 s - \frac{3}{2D} Q^2 sR \biggr] \nonumber \\
&+& q_{\alpha} q_{\beta} \frac{1}{D} \biggl[ R k^2 + l^4 s - \frac{3}{2D}
k^2 Q^2 sR \biggr] \nonumber \\
&-& (k_{\alpha} q_{\beta} + q_{\alpha} k_{\beta}) 
\frac{1}{D} \biggl[ R qk - l^2 Q^2 s - \frac{3}{2D} Q^2 qk sR \biggr] \biggl\}
\nonumber \\
&-& m^2 \biggl\{ g_{\alpha \beta} \biggl[ (k^2 - Q^2) + \frac{1}{2D} sR
\biggr] \nonumber \\ 
&+& k_{\alpha} k_{\beta} \frac{1}{D} \biggl[ R - Q^2 s + Q^4 - \frac{3}{2D}
Q^2 sR \biggr] \nonumber \\
&+& q_{\alpha} q_{\beta} \frac{1}{D} \biggl[ (l^2 + k^2)^2 - k^2 Q^2 
+ \frac{3}{2D} k^2 sR \biggr] \nonumber \\
&+& (k_{\alpha} q_{\beta} + q_{\alpha} k_{\beta}) 
\frac{1}{D} \biggl[ R - (l^2 + k^2)(qk - Q^2) + Q^2 qk - \frac{3}{2D} qk sR
\biggr] \biggl\}
\nonumber \\
&-& 2m^4 \biggl\{ g_{\alpha \beta} - k_{\alpha} k_{\beta} \frac{Q^2}{D}
+ q_{\alpha} q_{\beta} \frac{k^2}{D} - (k_{\alpha} q_{\beta} + 
q_{\alpha} k_{\beta}) \frac{qk}{D} \biggl\} \label{I.12}
\end{eqnarray}
and
\begin{eqnarray}
\tilde I_{\alpha \beta}^2 &=& \frac{s}{2} \biggl\{g_{\alpha \beta} 
\biggl[ k^2 + \frac{1}{2D} (Q^2 + k^2) R \biggr] \nonumber \\
&+& k_{\alpha} k_{\beta} \frac{1}{D} \biggl[ R - Q^2 (Q^2 + k^2) 
- \frac{3}{2D} Q^2 (Q^2 + k^2) R \biggr] \nonumber \\ 
&-& q_{\alpha} q_{\beta} \frac{1}{D} \biggl[ R - k^2 (Q^2 + k^2) 
- \frac{3}{2D} k^2 (Q^2 +k^2) R \biggr] \nonumber \\
&-& (k_{\alpha} q_{\beta} + q_{\alpha} k_{\beta}) \frac{1}{D} qk (Q^2 + k^2) 
\biggl[ 1 + \frac{3}{2D} R \biggr] \biggl\} \nonumber \\
&-& m^2 \biggl\{ -g_{\alpha \beta} \biggl[ Q^2 - \frac{1}{2D} sR \biggr] + 
k_{\alpha} k_{\beta} \frac{1}{D} \biggl[ R - 2Q^2 (qk - Q^2)
- \frac{3}{2D} Q^2 sR \biggr] \nonumber \\
&+& q_{\alpha} q_{\beta} \frac{1}{D} \biggl[ R + 2k^2 (qk - Q^2) 
+ \frac{3}{2D} k^2 sR \biggr] \nonumber \\
&+& (k_{\alpha} q_{\beta} + q_{\alpha} k_{\beta}) 
\frac{1}{D} \biggl[ R - (qk)^2 + 2qk Q^2 + Q^2 k^2 - \frac{3}{2D} qk sR
\biggr] \biggl\} \nonumber \\
&-& 2m^4 \biggl\{ g_{\alpha \beta} - k_{\alpha} k_{\beta} \frac{Q^2}{D}
+ q_{\alpha} q_{\beta} \frac{k^2}{D} - (k_{\alpha} q_{\beta} + 
q_{\alpha} k_{\beta}) \frac{qk}{D} \biggl\}. \label{I.13}
\end{eqnarray}

Equation~(\ref{I.11}) can be represented in the form
\begin{eqnarray}
C_{\alpha \beta}^{(a)} &=& A_{\alpha \beta} \frac{\alpha_s}
{4\pi} \frac{1}{\sqrt{D}} \int \limits_{l_-^2}^{l_+^2} dl^2
\Biggl\{ \frac{1}{l^2} \biggl[qk + \frac{1}{4D qk} (k^2 - Q^2) \Bigl(2(qk)^2
+ k^2Q^2 \Bigr)s \biggr] \nonumber \\
&+& \frac{1}{l^2} \biggl[ Q^2 + \frac{1}{2D} \Bigl(2(qk)^2 + k^2Q^2 \Bigr)s 
\biggr] \frac{m^2}{qk} - \frac{2}{l^2} \frac{m^4}{qk} \nonumber \\
&+& \frac{1}{2l^4} \biggl[ 1 - \frac{1}{2D} Q^2s \biggr] k^2Q^2
- \frac{1}{l^4} \biggl[ k^2 - Q^2 - \frac{1}{2D} k^2Q^2s \biggr]m^2 
\nonumber \\
&-& \frac{2}{l^4} m^4 - \frac{1}{2} \biggl[ 1 + \frac{1}{2D} k^2s \biggr]
\Biggr\} \nonumber \\
&+& B_{\alpha \beta} \frac{\alpha_s}{4\pi} \frac{1}{\sqrt{D}} 
\int \limits_{l_-^2}^{l_+^2} dl^2
\Biggl\{ \frac{1}{l^2} \biggl[qk + \frac{1}{4D qk} (k^2 - Q^2) \Bigl(2(qk)^2
- k^2Q^2 \Bigr)s \biggr] \nonumber \\
&+& \frac{1}{l^2} \biggl[ -4qk + 3Q^2 + \frac{3}{2D} \Bigl(2(qk)^2 + k^2Q^2 
\Bigr)s \biggr] \frac{m^2}{qk} - \frac{2}{l^2} \frac{m^4}{qk} \nonumber \\
&+& \frac{1}{2l^4} \biggl[ 1 - \frac{3}{2D} Q^2s \biggr] k^2Q^2
- \frac{1}{l^4} \biggl[ k^2 - Q^2 - \frac{3}{2D} k^2Q^2s \biggr]m^2 
\nonumber \\
&-& \frac{2}{l^4} m^4 - \frac{1}{2} \biggl[ 1 + \frac{3}{2D} k^2s \biggr]
\Biggr\}, \label{I.14}
\end{eqnarray}
where
\begin{eqnarray}
A_{\alpha \beta} &=& \biggl( g_{\alpha \beta} - \frac{k_{\alpha} 
k_{\beta}}{k^2} \biggl), \nonumber \\
B_{\alpha \beta} &=& (q_{\alpha} - k_{\alpha} \frac{qk}{k^2})
(q_{\beta} - k_{\beta} \frac{qk}{k^2}) \frac{k^2}{D}. \label{I.15}
\end{eqnarray}

From Eqs.~(\ref{I.14}) and (\ref{I.15}) one can see that 
$C_{\alpha \beta}^{(a)}$ obeys a condition
\begin{equation}
k^\alpha C_{\alpha \beta}^{(a)} = 0. \label{I.16}
\end{equation}
In deriving (\ref{I.14}) from (\ref{I.11})--(\ref{I.13}) we
took into account that $l_\pm^2 \rightarrow l_\mp^2$ at $l^2 
\rightarrow 2qk-l^2$, $l_\pm^2$ being defined by Eq.~(\ref{I.7}).

Note that the difference of the structure function under consideration,
$\Delta F_2$, is defined by $\Delta C_{\alpha \beta}^{(r)}$ ($r=a,b$).
The analysis shows that a leading contribution
into $\Delta C_{\alpha \beta}^{(r)}$ comes from the region $-k^2 \sim m^2
\ll Q^2$. Rejecting terms ${\rm O}(k^2/Q^2)$ and ${\rm O}(m^2/Q^2)$, 
we obtain the formula
\begin{eqnarray}
C_{\alpha \beta}^{(a)} &=& A_{\alpha \beta} \frac{\alpha_s}{4\pi} 
\int \limits_{l_-^2}^{l_+^2} dl^2 \biggl[ \frac{1}{l^2} 
\biggl( 1 - \frac{1}{2 (qk)^2} Q^2 s \biggr) + \frac{1}{2l^4} \frac{k^2Q^2}{qk}
\biggl( 1 - \frac{1}{2 (qk)^2} Q^2 s + 2 \frac{m^2}{k^2} \biggr) \nonumber \\
&-& \frac{1}{2(qk)} \biggr] 
+ B_{\alpha \beta} \frac{\alpha_s}{4\pi} 
\int \limits_{l_-^2}^{l_+^2} dl^2 \biggl[ \frac{1}{l^2} 
\biggl( 1 - \frac{1}{2 (qk)^2} Q^2 s \biggr) \nonumber \\
&+& \frac{1}{2l^4} \frac{k^2Q^2}{qk} \biggl( 1 - \frac{3}{2 (qk)^2} Q^2 s  
+ 2 \frac{m^2}{k^2} \biggr) - \frac{1}{2(qk)} \biggr]. \label{I.17}
\end{eqnarray}

Now let us consider the function $C_{\alpha \beta}^{(b)}$ (see 
Eqs.~(\ref{12})). The result of our calculations is of the form
\begin{eqnarray}
C_{\alpha \beta}^{(b)}(Q^2,k^2,qk) &=& \frac{3\alpha_s}{2\pi^2} 
\frac{Q^2}{D} \int d^4r \biggl[\frac{1}{l^4} J_{\alpha \beta}^1 (r,q,k;m^2) 
+ \frac{1}{l^2(2qk-l^2)} J_{\alpha \beta}^2 (r,q,k,;m^2)\biggr] 
\nonumber \\
&\times& \delta_+ ((q-r)^2 - m^2) \delta_+ ((k+r)^2 - m^2), \label{I.18}
\end{eqnarray}
where the contributions from ladder and crossed diagrams are, respectively,
\begin{eqnarray}
J_{\alpha \beta}^1 &=& g_{\alpha \beta} \frac{1}{2}
k^4 s + r_{\alpha} r_{\beta} 2[R - k^2 (2l^2 - Q^2) + k^2 s)] \nonumber \\
&+& k_{\alpha} k_{\beta} 2[R - k^2(l^2 - Q^2)] \nonumber \\
&+& (r_{\alpha} k_{\beta} + k_{\alpha} r_{\beta}) [2R - k^2 (3l^2 - 2Q^2) 
+ k^2 s] \nonumber \\
&+& (r_{\alpha} q_{\beta} + r_{\alpha} q_{\beta}) l^2 k^2 
+ (k_{\alpha} q_{\beta} + k_{\alpha} q_{\beta}) l^2 k^2 \label{I.19}
\end{eqnarray}
and
\begin{eqnarray}
J_{\alpha \beta}^2  = &-& g_{\alpha \beta} \frac{1}{2}
k^4 s + r_{\alpha} r_{\beta} 2(R - k^4) + k_{\alpha} k_{\beta} 
(l^2 - Q^2) k^2 \nonumber \\
&-& q_{\alpha} q_{\beta} (l^2 + k^2) k^2
+ (r_{\alpha} k_{\beta} + k_{\alpha} r_{\beta}) [R + k^2 (l^2 - k^2 - qk)]
\nonumber \\
&-& (r_{\alpha} q_{\beta} + r_{\alpha} q_{\beta}) [R - k^2 (l^2 + k^2 - qk)] 
\nonumber \\
&-& (k_{\alpha} q_{\beta} + k_{\alpha} q_{\beta}) \frac{1}{2} 
[2R - k^2 (k^2 - Q^2)]. \label{I.20}
\end{eqnarray}

Using formulae~(\ref{I.5}) and (\ref{I.8}) we obtain
\begin{eqnarray}
C_{\alpha \beta}^{(b)} = &-& A_{\alpha \beta} \frac{3\alpha_s}
{8\pi} \frac{Q^2}{D^{3/2}} \int \limits_{l_-^2}^{l_+^2} dl^2
\Biggl\{ \frac{1}{4l^2} \biggl[\frac{1}{qk}(k^2 - Q^2) \nonumber \\
&+& \frac{1}{D} \biggl(qk(3k^2 - 3Q^2 + 4qk) - 2k^2Q^2 \biggr) \biggr] k^2s
\nonumber \\
&+& \frac{1}{l^2} \biggl[ 2(k^2 + qk) + \frac{1}{qk} k^2(k^2 + Q^2)
\biggr] m^2 \nonumber \\
&-& \frac{1}{2l^4} \biggl[ 1 - \frac{1}{2D} Q^2s \biggr] k^4s
- \frac{1}{l^4} k^2s m^2 \nonumber \\
&+& l^2 \frac{1}{2D} (k^2 + qk)s - \frac{1}{2D} (k^2 + qk)
(k^2 + 2qk)s \Biggr\} \nonumber \\
&-& B_{\alpha \beta} \frac{3\alpha_s}
{8\pi} \frac{Q^2}{D^{3/2}} \int \limits_{l_-^2}^{l_+^2} dl^2
\Biggl\{ \frac{1}{4l^2} \biggl[- \frac{1}{qk}(k^2 - Q^2) \nonumber \\
&+& \frac{3}{D} \biggl(qk(3k^2 - 3Q^2 + 4qk) - 2k^2Q^2 \biggr) \biggr] k^2s
\nonumber \\
&+& \frac{1}{l^2} \biggl[ 2(k^2 + qk) + \frac{1}{qk} k^2(k^2 + Q^2)
\biggr] m^2 \nonumber \\
&-& \frac{1}{2l^4} \biggl[ 1 - \frac{3}{2D} Q^2s \biggr] k^4s
- \frac{1}{l^4} k^2s m^2 + l^2 \frac{3}{2D} (k^2 + qk)s \nonumber \\
&-& \frac{3}{2D} (k^2 + qk)(k^2 + 2qk)s + s \Biggr\}. \label{I.21}
\end{eqnarray}

At $-k^2 \sim m^2 \ll Q^2$ we get from (\ref{I.21})
\begin{eqnarray}
C_{\alpha \beta}^{(b)} = &-& A_{\alpha \beta} \frac{3\alpha_s}{8\pi}
\frac{Q^2}{(qk)^2} 
\int \limits_{l_-^2}^{l_+^2} dl^2 \biggl\{ \frac{1}{l^2} 
\biggl[ \frac{(qk - Q^2) s}{(qk)^2} + \frac{2m^2}{k^2} \biggl] k^2
\nonumber \\
&-& \frac{1}{2l^4} \frac{k^4 s}{qk}\biggl[1 - \frac{Q^2 s}{2(qk)^2} 
+ \frac{2m^2}{k^2}\biggl] - \frac{s}{qk} + l^2 \frac{s}{2(qk)^2}
\biggl\} \nonumber \\
&-& B_{\alpha \beta} \frac{3\alpha_s}{8\pi} \frac{Q^2}{(qk)^2} 
\int \limits_{l_-^2}^{l_+^2} dl^2 \biggl\{ \frac{1}{l^2} 
\biggl[\frac{(3qk - 2Q^2)s}{(qk)^2} + \frac{2m^2}{k^2} \biggl] k^2
\nonumber \\
&-& \frac{1}{2l^4} \frac{k^4 s}{qk}\biggl[1 - \frac{3Q^2 s}{2(qk)^2} 
+ \frac{2m^2}{k^2}\biggl] - \frac{2s}{qk} + l^2 \frac{3s}{2(qk)^2} 
\biggl\}. \label{I.22}
\end{eqnarray}

When calculating $C_{\alpha \beta}^{(b)}$ we used the replacement
\begin{equation}
\frac{p_{\alpha} p_{\beta}}{\sqrt{(pq)^2 + Q^2 M^2}} \rightarrow
\frac{k_{\alpha} k_{\beta}}{\sqrt{(qk)^2 + Q^2 k^2}}. \label{I.23}
\end{equation}
This results in power corrections that are, however, ignored everywhere 
in our consideration.
\vfill \eject

\def\theequation{II.\arabic{equation}}
\setcounter{equation}{0}

\section*{Appendix II}

In anology with quark distribution, scalar gluon distribution inside 
the nucleon, $D_g$, can be defined by considering deep inelastic scattering
with scalar gauge--invariant gluonic currents 
\begin{equation}
J(x) =\frac{1}{4} (G_{\mu \nu}^a(x))^2. \label{II.1}
\end{equation}

Let us define $D_g$ via a gluonic structure function $F$:
\begin{equation}
D(Q^2,x) = \frac{1}{x} F(Q^2,x) = \frac{1}{4\pi}{\rm Disc} T, \label{II.2}
\end{equation}
where
\begin{equation}
T = i\int d^4 \! z \exp (iqz) \langle p|TJ(z) J(0)|p \rangle. \label{II.3}
\end{equation}

Function $F$ is given by the formula
\begin{equation}
\frac{1}{x} F = \frac{1}{4\pi} \int \frac{d^4 \! k}{(2\pi)^4} \biggl[ 
\Pi_{\alpha \beta}(q,k) \frac{1}{k^4} \tilde \Gamma^{\alpha \beta} (k,p) 
\biggr]. \label{II.4}
\end{equation}
Here $\Pi_{\alpha \beta}$ is a gluonic partonometer
\begin{equation}
\Pi_{\alpha \beta} = - g_{\alpha \beta} 2\pi \delta ((q + k)^2) Q^2,
\label{II.5}
\end{equation}
while $\tilde \Gamma^{\alpha \beta}$ means an imaginary part of the virtual 
gluon--nucleon amplitude with tensor parts of the gluonic propagators 
$d_{\alpha \alpha'}$ included. So, we have by definition
\begin{equation}
\tilde \Gamma_{\alpha \beta} = d_{\alpha \alpha'}(k) d_{\beta \beta'}(k) 
\Gamma^{\alpha' \beta'}, \label{II.6}
\end{equation}
where the gluon distribution $\Gamma_{\alpha \beta}$ enters Eq.~(\ref{4}).

Tensor $\tilde \Gamma_{\alpha \beta}$ has the following Lorentz structure
\begin{equation}
\tilde \Gamma_{\alpha \beta} = d_{\alpha \beta}(p) \tilde \Gamma_1 
+ d_{\alpha \beta}(k) \tilde \Gamma_2 
+ \biggl(p_{\alpha} - k_{\alpha} \frac{pn}{kn} \biggr)
\biggl(p_{\beta} - k_{\beta} \frac{pn}{kn} \biggr) 
\frac{1}{k^2} \tilde \Gamma_3 
+ n_{\alpha} n_{\beta} \frac{k^2}{(kn)^2} \tilde \Gamma_4, \label{II.7}
\end{equation}
with $\tilde \Gamma_i=\tilde \Gamma_i(k^2,M^2,pk)$.
Then from Eqs.~(\ref{II.6}), (\ref{II.7}) and (\ref{13}) the
relation between $\tilde \Gamma_i$ and $\Gamma_i$ can easily be obtained. 
In particular, we have
\begin{equation}
\tilde \Gamma_3 = \Gamma_2 + \Gamma_4. \label{II.8}
\end{equation}

If we substitute (\ref{II.5}), (\ref{II.7}) into  (\ref{II.4}), we get
in LLA
\begin{equation}
D(Q^2,Q_0^2,x) = \int_{Q_0^2}^{Q^2} \frac{d k^2}{k^4} \int 
\frac{d^2 \! k_\bot}{2(2\pi)^4} \biggl[ \tilde \Gamma_1(k^2,k_\bot,x) 
+ \tilde \Gamma_2(k^2,k_\bot,x) + \frac{1 - x}{2x^2} 
\tilde \Gamma_3(k^2,k_\bot,x) \biggr]. \label{II.9}
\end{equation}

As can be seen, at small $x$ the gluon distribution~(\ref{II.9}) is
mainly defined by the invariant function $\tilde \Gamma_3$ (or, equivalently,
by the combination $\Gamma_2 + \Gamma_4$).
\vfill \eject

\vfill \eject

\section*{Figure Captions}

\noindent
{\bf Figs. 1a-1d:} \ $\Delta C(\eta, y)$ as a function of the variable $\eta$  
at several fixed values of $y$. \\

\noindent
{\bf Figs. 2a-2d:} \ $C(\eta, y)$ (continuous curves) and $\Delta C(\eta, y)$ 
                    (dashed curves) as functions of the \\ 
\hphantom{\bf Figs. 2a-2d:} \ 
                    variable $\eta$ ($\eta \geq 0$) at several fixed   
                    values of $y$.  


\begin{thebibliography}{9}
\bibitem{Data}
J. Chrin, Proc. of the 27th International Conference on High Energy
Physics, Glasgow,  1994 (Eds. P.J. Bussey and I.G. Knowles), 
p. 893;
A. De Angelis, Talk at the EPS Conference on High Energy Physics,
Brussels, 1995 (to appear in the Proceedings);
W.J. Metzger, {\it ibid.} 
\bibitem{Petrov-95}
V.A. Petrov and A.V. Kisselev, Z. Phys. C66 (1995) 453;
Nucl. Phys. B (Proc. Suppl.) 39B, C (1995) 364.
Early qualitative indications on the energy independence of the 
multiplicity difference were made by Yu. Dokshitzer, V.A. Khoze
and S.I. Troyan, J. Phys. G 17 (1991) 1481, 1602. 
\bibitem{Kisselev-95}
A.V. Kisselev and V.A. Petrov, Preprint IHEP 95-115, Protvino, 1995; 
Talk at the EPS Conference on High Energy Physics, 
Brussels, 1995 (to appear in the Proceedings).
\bibitem{Catani-91}
S. Catani, M. Ciafaloni and F. Hautmann, Nucl. Phys. B366 (1991) 135.
\bibitem{Kwiecinski-96}
J. Kwieci\'nski, A.D. Martin and P.J. Sutton, Preprint DTP/96/02,
Durham, 1996.
\bibitem{H1-96}
H1 Collaboration, S. Aid et al., Preprint DESY 96-039, Hamburg, 1996.
\bibitem{De Roeck-96}
A. De Roeck, Talk at the 2nd Krak\'ow Epiphany Conference, Krak\'ow, 1996
(to appear in the Proceedings).
\bibitem{Martin-94}
A.D. Martin, R.G. Roberts and W.J. Stirling, Phys. Rev. D50 (1994) 6734.
\end{thebibliography}
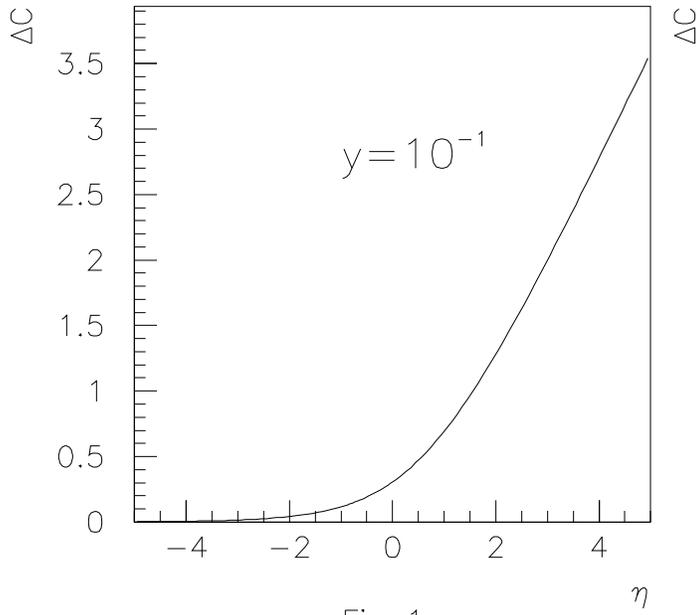
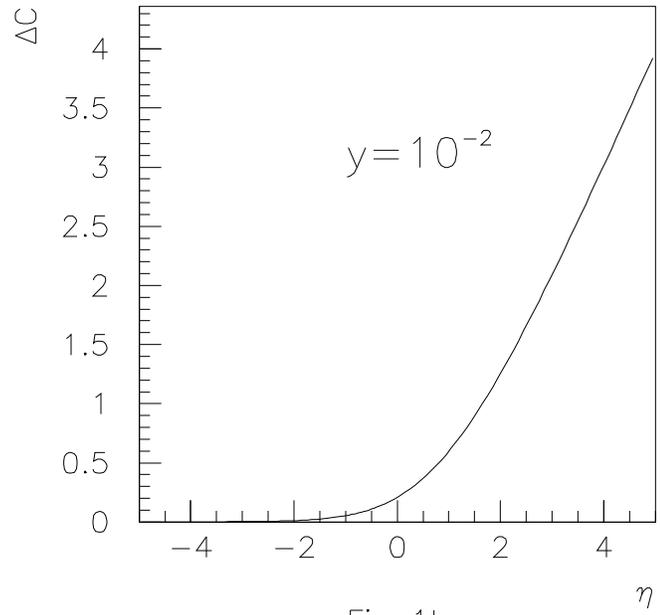
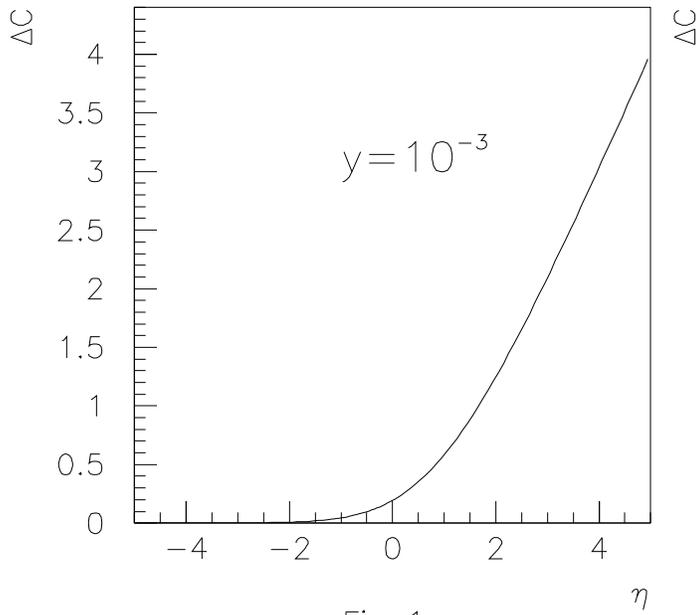
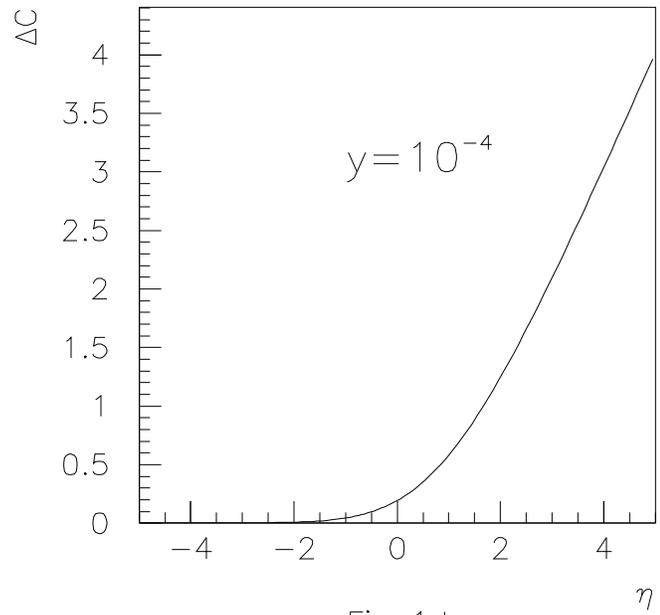
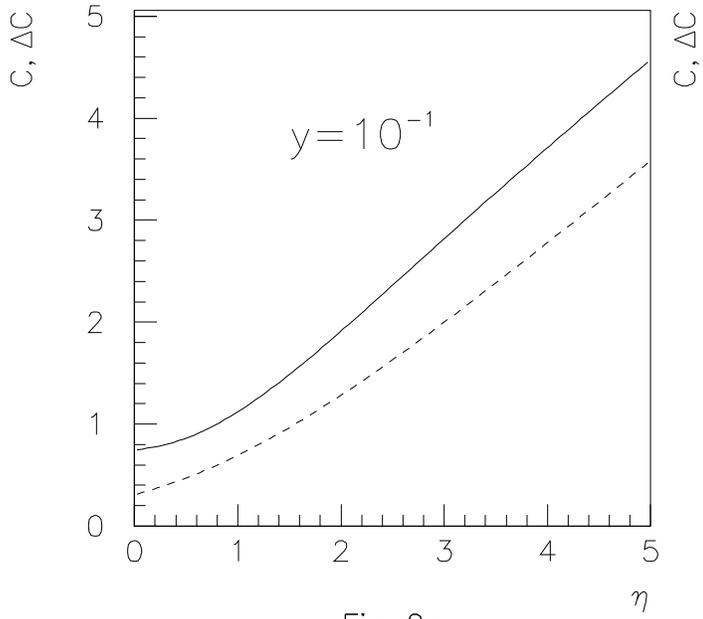
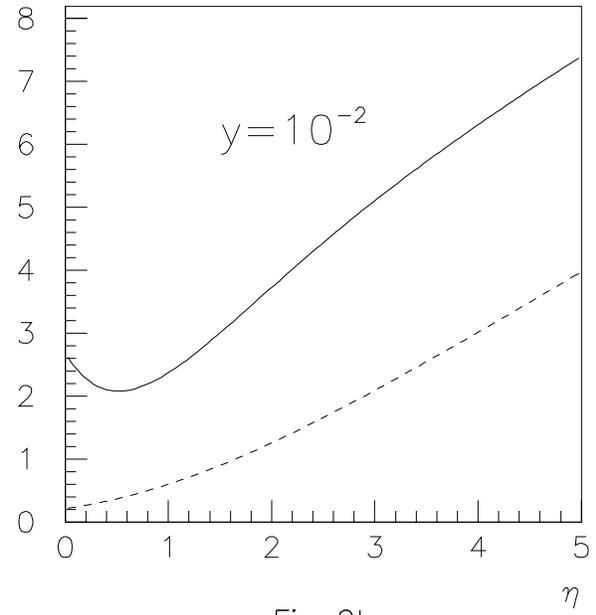
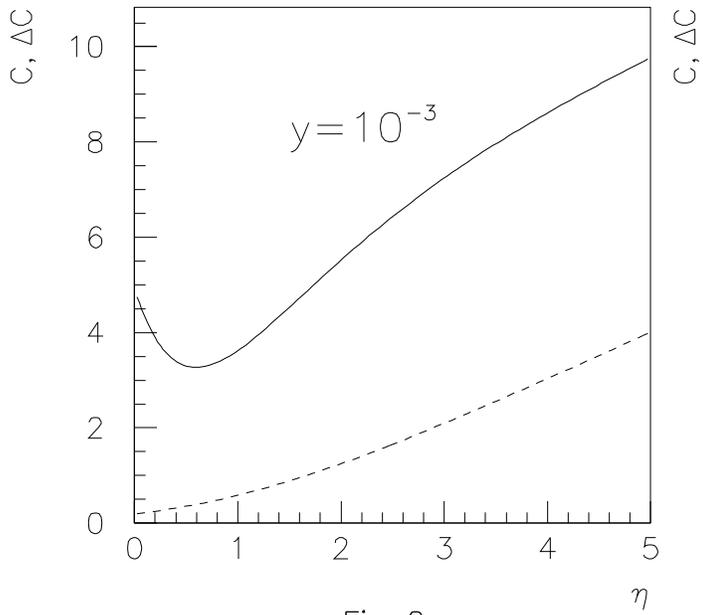
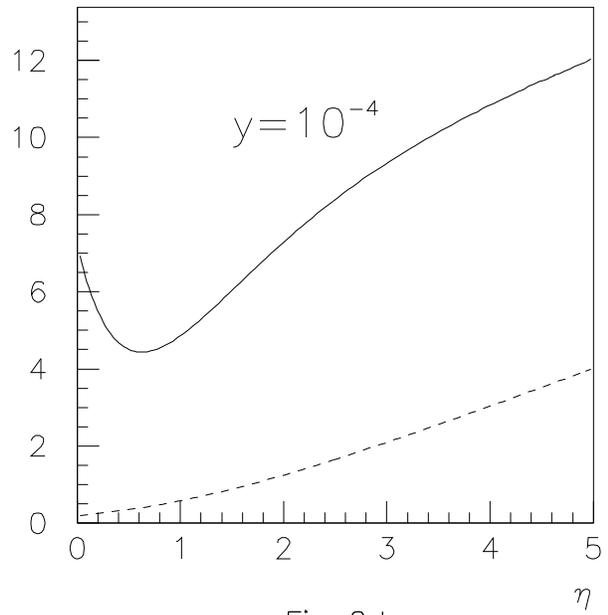
\end{document}